\documentstyle[twocolumn,aps]{revtex}

\begin{document}
\title{Chern-Simons Supergravities with Off-Shell Local Superalgebras \thanks{%
Talk presented at the Sixth Meeting on Quantum Mechanics of Fundamental
Systems: Black Holes and the Structure of the Universe, Santiago, August
1997.}}
\author{Ricardo Troncoso and Jorge Zanelli\thanks{%
John Simon Guggenheim fellow}}
\address{Centro de Estudios Cient\'{\i}ficos de Santiago, Casilla 16443, Santiago 9,
Chile\\
and \\
Departamento de F\'{\i}sica, Universidad de Santiago de Chile, Casilla 307,
Santiago 2, Chile.}
\maketitle

\begin{abstract}
A new family of supergravity theories in odd dimensions is presented. The
Lagrangian densities are Chern-Simons forms for the connection of a
supersymmetric extension of the anti-de Sitter algebra. The superalgebras
are the supersymmetric extensions of the AdS algebra for each dimension,
thus completing the analysis of van Holten and Van Proeyen, which was valid
for $N=1$ and for $D=2, 3, 4, mod \; 8$. The Chern-Simons form of the
Lagrangian ensures invariance under the gauge supergroup by construction
and, in particular, under local supersymmetry. Thus, unlike standard
supergravity, the local supersymmetry algebra closes off-shell and without
requiring auxiliary fields. The Lagrangian is explicitly given for $D=5,\;7$
and 11. In all cases the dynamical field content includes the vielbein ($%
e_{\mu }^{a}$), the spin connection ($\omega _{\mu }^{ab}$), $N$ gravitini ($%
\psi _{\mu }^{i}$), and some extra bosonic ``matter" fields which vary from
one dimension to another. The superalgebras fall into three families: $%
osp(m|N)$ for $D=2,3,4$, mod 8, $osp(N|m)$ for $D=6,7,8$, mod 8, and $%
su(m-2,2|N)$ for $D=5$ mod 4, with $m=2^{[D/2]}$. The possible connection
between the $D=11$ case and M-Theory is also discussed.
\end{abstract}


{\bf Abstract}


\section{Introduction}


A good part of the results presented in this lecture were also discussed in 
\cite{trz} and also presented at the January '98 meeting in Bariloche \cite
{trz'} --where the detailed construction of the superalgebra can be found--,
but it was at the meeting covered by these proceedings where these results
were first presented.

Three of the four fundamental forces of nature are consistently described by
Yang-Mills ({\bf YM}) quantum theories. Gravity, the fourth fundamental
interaction, resists quantization in spite of several decades of intensive
research in this direction. This is intriguing in view of the fact that
General Relativity ({\bf GR}) and YM theories have a deep geometrical
foundation: the gauge principle. How come two theories constructed on almost
the same mathematical basis produce such radically different physical
behaviours? What is the obstruction for the application of the methods of YM
quantum field theory to gravity? The final answer to these questions is
beyond the scope of this paper, however one can note a difference between YM
and GR which might turn out to be an important clue: YM theory is defined on
a fiber bundle, with the connection as the dynamical object, whereas the
dynamical fields of GR cannot be interpreted as components of a connection.
Therefore, gravitation does not lend itself naturally for a fiber bundle
interpretation.

The closest one could get to a connection formulation for GR is the Palatini
formalism, with the Hilbert action 
\begin{equation}
I[\omega ,e]=\int \epsilon _{abcd}R^{ab}\wedge e^{a}\wedge e^{b},
\label{hilbert}
\end{equation}
where $R^{ab}=d\omega ^{ab}+\omega _{c}^{a}\wedge \omega _{b}^{c}$ is the
curvature two-form, and $e^a$ is a local orthonormal frame. This action is
sometimes claimed to describe a gauge theory for local translations.
However, in our view this is a mistake. If $\omega$ and $e$ were the
components of the Poincar\'{e} connection, under local translations they
should transform as 
\begin{equation}
\delta \omega ^{ab}=0,\;\;\delta e^{a}=D\lambda ^{a}=d\lambda ^{a}+\omega
_{b}^{a}\wedge \lambda ^{b}.  \label{trans}
\end{equation}
Invariance of (\ref{hilbert}) under (\ref{trans})would require the
torsion-free condition, 
\begin{equation}
T^{a}=de^{a}+\omega _{b}^{a}\wedge e^{b}=0.  \label{T}
\end{equation}
This condition is an equation of motion for the action (\ref{hilbert}). This
means that the invariance of the action (\ref{hilbert}) under (\ref{trans})
could not result from the transformation properties of the fields alone, but
it would be a property of their dynamics as well. The torsion-free
condition, being one of the field equations, implies that local
translational invariance is at best an {\em on-shell} symmetry, which would
probably not survive quantization.

The contradiction stems from the identification between local translations
in the base manifold (diffeomorphisms) 
\begin{equation}
x^{\mu }\rightarrow x^{\prime }{}^{\mu }=x^{\mu }+\zeta ^{\mu }(x),
\label{diffeos}
\end{equation}
--which is a genuine invariance of the action (\ref{hilbert})--, and local
translations in the tangent space (\ref{trans}).

Since the invariance of the Hilbert action under general coordinate
transformations (\ref{diffeos}) is reflected in the closure of the
first-class hamiltonian constraints in the Dirac formalism, one could try to
push the analogy between the Hamiltonian constraints $H_{\mu}$ and the
generators of a gauge algebra. However, the fact that the constraint algebra
requires structure {\em functions}, which depend on the dynamical fields, is
another indication that the generators of diffeomorphism invariance of the
theory do not form a Lie algebra but an open algebra (see, e. g., \cite{MH}).

More precisely, the subalgebra of spatial diffeomorphisms {\em is} a genuine
Lie algebra in the sense that its structure constants are independent of the
dynamical fields of gravitation, 
\begin{equation}
[H_i, H_j^{\prime}] \sim H_j^{\prime}\delta_{|i} - H_i^{\prime}\delta_{|j}.
\label{hi}
\end{equation}
In contrast, the generators of timelike diffeomorphisms form an open
algebra, 
\begin{equation}
[H_{\perp}, H_{\perp}^{\prime}] \sim g^{ij}H_j^{\prime}\delta_{|i}.
\label{hperp}
\end{equation}

This comment is particularly relevant in a CHern-Simons theory, where
spatial diffeomorphisms are always part of the true gauge symmetries of the
theory. The generators of timelike displacements ($H_{\perp}$), on the other
hand, are combinations of the internal gauge generators and the generators
of spatial diffeomorphism, and therefore do not generate independent
symmetries \cite{bgh}.\newline

{\bf Higher D} The minimal requirements for a consistent theory which
includes gravity in any dimension are: general covariance and second order
field equations for the metric. For $D>4$ the most general action for
gravity satisfying this criterion is a polynomial of degree $[D/2]$ in the
curvature, first discussed by Lanczos for $D=5$ \cite{lanczos} and, in
general, by Lovelock \cite{lovelock,zumino}.\newline

{\bf First order theory}

If the theory contains spinors that couple to gravity, it is necessary to
decouple the affine and metric properties of spacetime. A metric formulation
is sufficient for spinless point particles and fields because they only
couple to the symmetric part of the affine connection, while a spinning
particle can ``feel'' the torsion of spacetime. Thus, it is reasonable to
look for a formulation of gravity in which the spin connection ($\omega_{\mu
}^{ab}$) and the vielbein ($e_{\mu }^{a}$) are dynamically independent
fields, with curvature and torsion standing on a similar footing. Thus, the
most general gravitational Lagrangian would be of the general form $%
L=L(\omega, e)$ \cite{mz}.

Allowing an independent spin connection in four dimensions does not modify
the standard picture in practice because any occurrence of torsion in the
action leaves the classical dynamics essentially intact. In higher
dimensions, however, theories that include torsion can be dynamically quite
different from their torsion-free counterparts.

As we shall see below, the dynamical independence of $\omega ^{ab}$ and $%
e^{a}$ also allows defining these gravitation theories in $2n+1$ dimensions
on a fiber bundle structure as a Yang-Mills theory, a feature that is not
shared by General Relativity except in three dimensions.


\section{Supergravity}


For some time it was hoped that the nonrenormalizability of GR could be
cured by supersymmetry. However, the initial glamour of supergravity ({\bf %
SUGRA}) as a mechanism for taming the wild ultraviolet divergences of pure
gravity, was eventually spoiled by the realization that it too would lead to
a nonrenormalizable answer \cite{pkt84}. Again, one can see that SUGRA is
not a gauge theory either in the sense of a fiber bundle, and that the local
symmetry algebra closes naturally only on shell. The algebra can be made to
close off shell at the cost of introducing auxiliary fields, but they are
not guaranteed to exist for all $D$ and $N$ \cite{tr}.

Whether the lack of fiber bundle structure is the ultimate reason for the
nonrenormalizability of gravity remains to be proven. However, it is
certainly true that if GR could be formulated as a gauge theory, the chances
for proving its renormalizability would clearly grow.

In three spacetime dimensions both GR and SUGRA define renormalizable
quantum theories. It is strongly suggestive that precisely in 2+1 dimensions
both theories can also be formulated as gauge theories on a fiber bundle 
\cite{witten}. It might seem that the exact solvability miracle was due to
the absence of propagating degrees of freedom in three-dimensional gravity,
but the power counting renormalizability argument rests on the fiber bundle
structure of the Chern-Simons form of those systems.

There are other known examples of gravitation theories in odd dimensions
which are genuine (off-shell) gauge theories for the anti-de Sitter ({\bf AdS%
}) or Poincar\'{e} groups \cite{chamslett,chamseddine,btz,z}. These
theories, as well as their supersymmetric extensions have propagating
degrees of freedom \cite{bgh} and are CS systems for the corresponding
groups as shown in \cite{btrz}.


\subsection{From Rigid Supersymmetry to Supergravity}


Rigid SUSY can be understood as an extension of the Poincar\'e algebra by
including supercharges which are the ``square roots" of the generators of
rigid translations, $\{\bar{Q},Q\} \sim \Gamma\cdot \mbox{P}$. The basic
strategy to generalize this idea to local SUSY was to substitute the
momentum P$_{\mu} = i\partial_{\mu}$ by the generators of diffeomorphisms, $%
{\cal H}$, and relate them to the supercharges by $\{\bar{Q},Q\} \sim
\Gamma\cdot{\cal H}$. The resulting theory has on-shell local supersymmetry
algebra \cite{pvn}.

An alternative point of view --which is the one we advocate here-- would be
to construct the supersymmetry on the tangent space and not on the base
manifold. This approach is more natural if one recalls that spinors provide
a basis of irreducible representations for $SO(N)$, and not for $GL(N)$.
Thus, spinors are naturally defined relative to a local frame on the tangent
space rather than in the coordinate basis. The basic point is to reproduce
the 2+1 ``miracle" in higher dimensions. This idea has been successfully
applied by Chamseddine in five dimensions\cite{chamseddine}, and by us for
pure gravity \cite{btz,z} and in supergravity \cite{trz,btrz}. The SUGRA
construction has been carried out for spacetimes whose tangent space has AdS
symmetry \cite{trz}, and for its Poincar\'e contraction in \cite{btrz}.

In \cite{btrz}, a family of theories in odd dimensions, invariant under the
supertranslation algebra whose bosonic sector contains the Poincar\'{e}
generators was presented. The anticommutator of the supersymmetry generators
gives a translation plus a tensor ``central" extension, 
\begin{equation}
\{Q^{\alpha },\bar{Q}_{\beta }\}=-i(\Gamma ^{a})_{\beta }^{\alpha}P_{a}
-i(\Gamma^{abcde})_{\beta }^{\alpha }Z_{abcde},  \label{supertrans}
\end{equation}
The commutators of $Q,\bar{Q}$ and $Z$ with the Lorentz generators can be
read off from their tensorial character. All the remaining commutators
vanish. This algebra is the continuation to all odd-dimensional spacetimes
of the $D=10$ superalgebra of van Holten and Van Proeyen \cite{vV}, and
yields supersymmetric theories with off-shell Poincar\'{e} superalgebra. The
existence of these theories suggests that there should be similar
supergravities based on the AdS algebra.


\subsection{Assumptions of Standard Supergravity}


Three implicit assumptions are usually made in the construction of standard
SUGRA:

{\bf (i)} The fermionic and bosonic fields in the Lagrangian should come in
combinations such that their propagating degrees of freedom are equal in
number. This is usually achieved by adding to the graviton and the gravitini
a number of lower spin fields ($s<3/2$)\cite{pvn}. This matching, however,
is not necessarily true in AdS space, nor in Minkowski space if a different
representation of the Poincar\'e group (e.g., the adjoint representation) is
used \cite{soh}.

The other two assumptions concern the purely gravitational sector. They are
as old as General Relativity itself and are dictated by economy: {\bf (ii)}
gravitons are described by the Hilbert action (plus a possible cosmological
constant), and, {\bf (iii)} the spin connection and the vielbein are not
independent fields but are related through the torsion equation. The fact
that the supergravity generators do not form a closed off-shell algebra can
be traced back to these asumptions.

The procedure behind {\bf (i)} is tightly linked to the idea that the fields
should be in a {\em vector} representation of the Poincar\'{e} group \cite
{soh} and that the kinetic terms and couplings are such that the counting of
degrees of freedom works like in a minimally coupled gauge theory. This
assumption comes from the interpretation of supersymmetric states as
represented by the in- and out- plane waves in an asymptotically free,
weakly interacting theory in a minkowskian background. These conditions are
not necessarily met by a CS theory in an asymptotically AdS background.
Apart from the difference in background, which requires a careful treatment
of the unitary irreducible representations of the asymptotic symmetries \cite
{Gunaydin}, the counting of degrees of freedom in CS theories is completely
different from the one for the same connection one-forms in a YM theory.


\section{Lanczos--Lovelock Gravity}


\subsection{Lagrangian}


For $D>4$, assumption {\bf (ii)} is an unnecessary restriction on the
available theories of gravitation. In fact, as mentioned above, the most
general action for gravity --generally covariant and with second order field
equations for the metric-- is the Lanczos-Lovelock Lagrangian ({\bf LL}).
The LL Lagrangian in a D-dimensional Riemannian manifold can be defined in
at least four ways:

({\bf a}) As the most general invariant constructed from the metric and
curvature leading to second order field equations for the metric \cite
{lanczos,lovelock,zumino}.

({\bf b}) As the most general $D$-form invariant under local Lorentz
transformations, constructed with the vielbein, the spin connection, and
their exterior derivatives, without using the Hogde dual ($*$) \cite{Regge}.

({\bf c}) As a linear combination of the dimensional continuation of all the
Euler classes of dimension $2p<D$.\cite{zumino,tz}

({\bf d}) As the most general low energy effective gravitational theory that
can be obtained from string theory \cite{zwiebach}.

Definition ({\bf a}) was historically the first. It is appropriate for the
metric formulation and assumes vanishing torsion. Definition ({\bf b}) is
slightly more general than the first and allows for a coordinate-independent
first-order formulation, and even allows torsion-dependent terms in the
action \cite{mz}. As a consequence of ({\bf b}), the field configurations
that extremize the action obey first order equations for $\omega$ and $e$.
Assertion ({\bf c}) gives directly the Lanczos--Lovelock solution as a
polynomial of degree $[D/2]$ in the curvature of the form

\begin{equation}
I_{G}=\int \sum_{p=0}^{[D/2]}\alpha _{p}L^{p},  \label{Lovaction}
\end{equation}
where $\alpha _{p}$ are arbitrary constants and\footnote{%
For even and odd dimensions the same expression (\ref{Lovlag}) can be used,
but for odd $D$, Chern-Simons forms for the Lorentz connection could also be
included (this point is discussed below).}

\begin{equation}
L_{G}^{p}=\epsilon _{a_{1}\cdots a_{D}}R^{a_{1}a_{2}}\cdots
R^{a_{2p-1}a_{2p}}e^{a_{2p+1}}\cdots e^{a_{D}},  \label{Lovlag}
\end{equation}
where wedge product of forms is understood throughout.

Statement ({\bf d}) reflects the empirical observation that the vanishing of
the superstring $\beta$-function in $D=10$ gives rise to an effective
Lagrangian of the form (\ref{Lovlag}) \cite{zwiebach}.

In even dimensions, the last term in the sum is the Euler character, which
does not contribute to the equations of motion. However, in the quantum
theory, this term in the partition function would assign different weights
to nonhomeomorphic geometries.

The large number of dimensionful constants $\alpha _{p}$ in the LL theory
contrasts with the two constants of the EH action ($G$ and $\Lambda $) \cite
{bd,jjg,btz}. This feature could be seem as an indication that
renormalizability would be even more remote for the LL theory than in
ordinary gravity. However, this is not necessarily so. There are some very
special choices of $\alpha _{p}$ such that the theory becomes invariant
under a larger gauge group in odd spacetime dimensions, which could actually
improve renormalizability \cite{witten,z}.


\subsection{Equations}


Consider the Lovelock action (\ref{Lovaction}), viewed as a functional of
the spin connection and the vielbein, 
\begin{equation}
I_{LL}=I_{LL}\left[ \omega ^{ab},e^{a}\right] .  \label{firstL}
\end{equation}

Varying with respect to the vielbein, the generalized Einstein equations are
obtained, 
\begin{eqnarray}
\sum_{p=0}^{n-1}\alpha_{p}(D-2p)\epsilon_{a_{1}\cdots a_{D}}R^{a_{1}a_{2}}
\cdots R^{a_{2p-1}a_{2p}}\times & &  \nonumber \\
e^{a_{2p+1}} \cdots e^{a_{D-1}} & = & 0.  \label{E-L}
\end{eqnarray}
Varying with respect to the spin connection, the torsion equations are
found, 
\begin{eqnarray}
\sum_{p=0}^{n-1}\alpha _{p}p(D-2p)\epsilon _{aba_{3}\cdots
a_{D}}R^{a_{3}a_{4}}\cdots R^{a_{2p-1}a_{2p}} \times &&  \nonumber \\
e^{a_{2p+1}}\cdots e^{a_{D-1}}T^{a_{D}} &=&0.  \label{Tor}
\end{eqnarray}

The presence of the arbitrary coefficients $\alpha _{p}$ in the action
implies that static, spherically symmetric Schwarzschild-like solutions
possess a large number of horizons \cite{wheeler}, and time-dependent
solutions have an unpredictable evolution \cite{tz,htz}. However, as shown
below, for a particular choice of the constants $\alpha_{p}$ the dynamics is
significantly better behaved.

Additional terms containing torsion explicitly can be included in the
action. It can be shown, however, that the presence of torsional terms in
the Lagrangian does not change the degrees of freedom of gravity in four
dimensions. Indeed, the matter-free theory with torsion terms is
indistinguishable (at least classically) from GR, \cite{hojman}. However, in
higher dimensions, the situation is completely different \cite{mz}.


\subsection{The vanishing of Classical Torsion}


Obviously $T^{a}=0$ solves (\ref{Tor}). However, for $D>4$ this equation
does not imply vanishing torsion in general. In fact, there are choices of
the coefficients $\alpha _{p}$ and configurations of $\omega^{ab}$, $e^{a}$
such that $T^{a}$ is completely arbitrary. On the other hand, as already
mentioned, the torsion-free postulate is at best a good description of the
classical dynamics only. Thus, an off-shell treatment of gravity should
allow for dynamical torsion even in four dimensions. In the first order
formulation, the theory has second class constraints due to the presence of
a large number of ``coordinates" which are actually ``momenta" \cite
{mauricio}, thus complicating the dynamical analysis of the theory.

On the other hand, if torsion is assumed to vanish, $\omega $ could be
solved as a function of $e^{-1}$ and its first derivatives, but this would
restrict the validity of the approach to nonsingular configurations for
which $\det(e_{\mu}^a)\neq 0$. In this framework, the theory has no second
class constraints and the number of degrees of freedom is the same as in the
Einstein-Hilbert theory, namely $\frac{D(D-3)}{2}$ \cite{tz}.


\subsection{Dynamics and Degrees of Freedom}


Imposing $T^a=0$ from the start, the action is $I_=I_{LL}[e^a,\omega(e)]$
and varying respect to $e$, the ``1.5 order formalism'' \cite{pvn} is
obtained,

\begin{equation}
\delta I_=\frac{\delta I_{LL}}{\delta e^a}\delta e^a+\frac{\delta I_{LL}} {%
\delta \omega ^{bc}}\frac{\delta \omega ^{bc}}{\delta e^a}\delta e^a.
\end{equation}

Assuming $\frac{\delta I_{LL}}{\delta \omega^{bc}}=0$ the equations of
motion consist of the Einstein equations (\ref{E-L}), defined on a
restricted configuration space.

For $D\leq 4$, $T^{a}=0$ is the unique solution of eqn.(\ref{Tor}). In those
dimensions, the different variational principles (first-, second- and 1.5-th
order) are classically equivalent in the absence of sources. On the
contrary, for $D>4$, $T^{a}=0$ is not logically necessary and is therefore
unjustified.

The LL--Lagrangians (\ref{Lovlag}) include the Einstein-Hilbert ({\bf EH})
theory as a particular case, but they are dynamically very different in
general. The classical solutions of the LL theory are not perturbatively
related to those of the Einstein theory. For instance, it was observed that
the time evolution of the classical solutions in the LL theory starting from
a generic initial state can be unpredictable, whereas the EH theory defines
a well-posed Cauchy problem.

It can also be seen that even for some simple minisuperspace models, the
dynamics could become quite messy because the equations of motion are not
deterministic in the classical sense, due to the vanishing of some
eigenvalues of the Hessian matrix on critical surfaces in phase space \cite
{tz,htz}.


\subsection{Choice of Coefficients}


At least for some simple minisuperspace geometries the indeterminate
classical evolution can be avoided if the coefficients are chosen so that
the Lagrangian is based on the connection for the AdS group, 
\begin{equation}
\alpha _{p}l^{D-2p}=\left\{
\begin{array}{ll}
(D-2p)^{-1}\left(
\begin{array}{c}
n-1 \\ 
p
\end{array}
\right) , & D=2n-1 \\ 
\left(
\begin{array}{c}
n \\ 
p
\end{array}
\right) , & D=2n.
\end{array}
\right.  \label{BI/CS}
\end{equation}
This corresponds to the Born-Infeld theory in even dimensions \cite{jjg},
and to the AdS Chern-Simons theory in odd dimensions \cite
{btz,chamslett,chamseddine},


\subsubsection{$D=2n$: Born-Infeld Gravity}


In even dimensions the choice (\ref{BI/CS}) gives rise to a Lagrangian of
the form 
\begin{equation}
L = \kappa \epsilon_{a_1\cdots a_D} (R^{a_1 a_2} +\frac{e^{a_1} e^{a_2}}{l^2}%
) \cdots (R^{a_{D-1} a_D} +\frac{e^{a_{D-1}} e^{a_D}}{l^2}).  \label{BI}
\end{equation}
This is the Pfaffian of the two--form $R^{a b} +\frac{1}{l^2} e^a e^b$, and,
in this sense it can be written in the Born-Infeld-like form, 
\begin{equation}
L= \kappa \sqrt{\mbox{det}(R^{a b} +\frac{1}{l^2} e^a e^b)}.  \label{BI'}
\end{equation}

The combinations $R^{ab}+\frac{1}{l^2}e^a e^b$ are the components of the AdS
curvature (c.f.(\ref{R}) below). This seems to suggest that the system might
be naturally described in terms of an AdS connection \cite{freund}. However,
this is not the case: In even dimensions, the Lagrangian (\ref{BI}) is
invariant under local Lorentz transformations and not under the entire AdS
group. As will be shown below, it is possible, in odd dimensions, to
construct gauge invariant theories of gravity under the full AdS group.


\subsubsection{$D=2n-1$: AdS Gauge Gravity}


The odd-dimensional case was discussed in \cite{chamslett,chamseddine}, and
later also in \cite{btz}. Consider the action (\ref{Lovaction}) with the
choice given by (\ref{BI/CS}) for $D=2n-1$. The constant parameter $l$ has
dimensions of length and its purpose is to render the action dimensionless.
This also allows the interpretation of $\omega $ and $e$ as components of
the AdS connection \cite{jjg}, $A=\frac{1}{2}\omega^{ab}J_{ab}+
e^{a}J_{aD+1}=\frac{1}{2} W^{AB}J_{AB}$, where 
\begin{equation}
W^{AB}=\left[ 
\begin{array}{cc}
\omega ^{ab} & e^{a}/l \\ 
-e^{b}/l & 0
\end{array}
\right] ,\;A,B=1,...D+1.  \label{w}
\end{equation}

The resulting Lagrangian is the Euler-CS form. Its exterior derivative is
the Euler form in $2n$ dimensions, 
\begin{eqnarray}
dL_{G\;2n-1}^{AdS} &=& \kappa \epsilon _{A_{1}\cdots
A_{2n}}R^{A_{1}A_{2}}\cdots R^{A_{2n-1}A_{2n}} \\
&=& \kappa{\cal E}_{2n},  \nonumber  \label{E}
\end{eqnarray}
where $R^{AB} = dW^{AB}+W_{C}^{A}W^{CB}$ is the AdS curvature, which
contains the Riemann and torsion tensors, 
\begin{equation}
R^{AB} = \left[ 
\begin{array}{cc}
R^{ab}+\frac{1}{l^{2}}e^{a}e^{b} & T^{a}/l \\ 
-T^{b}/l & 0
\end{array}
\right].  \label{R}
\end{equation}
The constant $\kappa$ is quantized \cite{z} (in the following we will set $%
\kappa =l=1$).

In general, a Chern-Simons Lagrangian in $2n-1$ dimensions is defined by the
condition that its exterior derivative be an invariant homogeneous
polynomial of degree $n$ in the curvature, that is, a characteristic class.
In the case above, (\ref{E}) defines the CS form for the Euler class $2n$%
-form.

A generic CS Lagrangian in $2n-1$ dimensions for a Lie algebra $g$ can be
defined by 
\begin{equation}
dL_{2n-1}^{g}=\left\langle \mbox{{\bf F}}^{n}\right\rangle ,  \label{F^n}
\end{equation}
where $\left\langle \ \right\rangle $ stands for a multilinear function in
the Lie algebra $g$, invariant under cyclic permutations such as Tr, for an
ordinary Lie algebra, or STr, in the case of a superalgebra. In the case
above, the only nonvanishing brackets in the algebra are 
\begin{equation}
\left\langle J_{A_1 A_2},\cdots, J_{A_{D-1}A_D} \right\rangle =\epsilon_{A_1
\cdots A_D}.  \label{Bracket}
\end{equation}


\subsubsection{$D=2n-1$: Poincar\'{e} Gauge Gravity}


Starting from the AdS theory (\ref{E}) in odd dimensions, a Wigner-
In\"{o}n\"{u} contraction deforms the AdS algebra into the Poincar\'e one.
The same result is also obtained choosing $\alpha_p=\delta^n_p$. Then, the
Lagrangian (\ref{Lovaction}) becomes:

\begin{equation}
L_{G}^{P}=\epsilon _{a_1\cdots a_D}R^{a_1 a_2} \cdots R^{a_{D-2}a_{D-1}}
e^{a_{D}}.  \label{14}
\end{equation}

In this way the local symmetry group of (\ref{Lovaction}) is extended from
Lorentz ($SO(D-1,1)$) to Poincar\'{e} ($ISO(D-1,1)$). Analogously to the
anti-de Sitter case, one can see that the action depends on the Poincar\'{e}
connection: ${\bf A} = e^{a}P_{a}+\frac{1}{2}\omega^{ab}J_{ab}$. It is
straightforward to verify the invariance of the action under local
translations, 
\begin{equation}
\delta e^{a}=D\lambda^{a},\mbox{\hspace{.2cm}}\delta \omega^{ab}=0,
\label{translations}
\end{equation}
Here $D$ stands for covariant derivative in the Lorentz connection. If $%
\lambda$ is the Lie algebra-valued zero-form, $\lambda=\lambda^{a}P_{a}$,
the transformations $($\ref{translations}$)$ are read from the general gauge
transformation for the connection, $\delta {\bf A}=\nabla \lambda $, where $%
\nabla $ is the covariant derivative in the Poincar\'{e} connection.

Moreover, the Lagrangian (\ref{14}) is a Chern-Simons form. Indeed, with the
curvature for the Poincar\'e algebra, {\bf F}$=d${\bf A} +{\bf A}$%
\mbox{\tiny $\wedge$}${\bf A} =$\frac{1}{2}R^{ab}J_{ab}+T^{a}P_{a}$, $%
L_{G}^{P}$ satisfies

\begin{equation}
dL_{G}^{P} = \left\langle \mbox{{\bf F}}^{n+1}\right\rangle,
\label{PPoincare}
\end{equation}
where the only nonvanishing components in the bracket are 
\begin{equation}
\left\langle J_{a_{1}a_{2}},\cdots,
J_{a_{D-2}a_{D-1}},P_{a_{D}}\right\rangle =\epsilon _{a_1 \cdots a_D}.
\label{PBracket}
\end{equation}

Thus, the Chern character for the Poincar\'e group is written in terms of
the Riemman curvature and the torsion as

\begin{equation}
\left\langle \mbox{{\bf F}}^3\right\rangle=\epsilon_{a_1 \cdots a_D}
R^{a_1a_2} \cdots R^{a_{D-2}a_{D-1}} T^{a_D}.
\end{equation}

The simplest example of this is ordinary gravity in 2+1 dimensions, where
the Einstein-Hilbert action with cosmological constant is a genuine {\em %
gauge} theory of the AdS group, while for zero cosmological constant it is
invariant under {\em local} Poincar\'e transformations. Although this gauge
invariance of 2+1 gravity is not always emphasized, it lies at the heart of
the proof of integrability of the theory \cite{witten}.


\section{AdS Gauge Gravity}


As shown above, the LL action assumes spacetime to be a Riemannian,
torsion-free, manifold. That assumption is justified {\em a posteriori} by
the observation that $T^a=0$ is always a solution of the classical
equations, and means that $e$ and $\omega$ are not dynamically independent.
This is the essence of the second order or metric approach to GR, in which
distance and parallel transport are not independent notions, but are related
through the Christoffel symbol. There is no fundamental justification for
this assumption and this was the issue of the historic discussion between
Einstein and Cartan \cite{E-C}.

In four dimensions, the equation $T^{a}=0$ is algebraic and could in
principle be solved for $\omega$ in terms of the remaining fields. However,
for $D>5$, CS gravity has more degrees of freedom than those encountered in
the corresponding second order formulation \cite{bgh}. This means that the
CS gravity action has propagating degrees of freedom for the spin
connection. This is a compelling argument to seriously consider the
possibility of introducing torsion terms in the Lagrangian from the start.

Another consequence of imposing a dynamical dependence between $\omega $ and 
$e$ through the torsion-free condition is that it spoils the possibility of
interpreting the local translational invariance as a gauge symmetry of the
action. Consider the action of the Poincar\'{e} group on the fields as given
by $($\ref{translations}$)$; taking $T^a\equiv 0$ implies 
\begin{equation}
\delta \omega^{ab}=\frac{\delta \omega ^{ab}}{\delta e^{c}}\delta e^{c}\neq
0,
\end{equation}
which would be inconsistent with the transformation of the fields under
local translations (\ref{trans}). Thus, the spin connection and the vielbein
--the soldering between the base manifold and the tangent space-- cannot be
identified as the compensating fields for local Lorentz rotations and
translations, respectively.

In our construction $\omega $ and $e$ are assumed to be dynamically
independent and thus torsion necessarily contains propagating degrees of
freedom, represented by the contorsion tensor $k_{\mu}^{ab}:=
\omega_{\mu}^{ab}-\bar{\omega}_{\mu}^{ab}(e,...)$, where $\bar{\omega}$ is
the spin connection which solves the (algebraic) torsion equation in terms
of the remaining fields.

The generalization of the Lovelock theory to include torsion explicitly can
be obtained assuming definition {\bf (b)}. This is a cumbersome problem due
to the lack of a simple algorithm to classify all possible invariants
constructed from $e^{a},R^{ab}$ and $T^{a}$. In Ref. \cite{mz} a useful
``recipe'' to generate all those invariants is given.


\subsection{The Two Families of AdS Theories}


Similarly to the theory discussed in section III, the torsional additions to
the Lagrangian bring in a number of arbitrary dimensionful coefficients $%
\beta_k$, analogous to the $\alpha_p$'s. Also in this case, one can try
choosing the $\beta$'s in such a way as to enlarge the local Lorentz
invariance into an AdS gauge symmetry. If no additional structure (e.g.,
inverse metric, Hodge-$*$, etc.) is assumed, AdS invariants can only be
produced in dimensions $4k$ and $4k-1$.

The proof of this claim is as follows: invariance under AdS requires that
the D-form be at least Lorentz invariant. Then, in order for these scalars
to be invariant under AdS as well, it is necessary and sufficient that they
be expressible in terms of the AdS connection (\ref{w}). As is well-known
(see, e.g., \cite{nakahara}), in even dimensions, the only $D$-form
invariant under $SO(N)$ constructed according to the recipe mentioned above
are\footnote{%
For simplicity we will not always distinguish between different signatures.
Thus, if no confussion can occur, the AdS group in $D$ dimensions will also
be denoted as $SO(D+1)$. The de Sitter case can be obtained replacing $%
\alpha_p$ by $(-1)^p\alpha_p$ in (\ref{BI/CS}).} the Euler character (for $%
N=D$), and the Chern characters (for any $N$). Thus, the only AdS invariant $%
D$-forms are the Euler class, and linear conbinations of products of the
type 
\begin{equation}
P_{r_1 \cdots r_s} = c_{r_1} \cdots c_{r_s},  \label{chern}
\end{equation}
with $2(r_1+r_2+\cdots+ r_s) = D$, where 
\begin{equation}
c_r = \mbox{Tr({\bf F}}^r),
\end{equation}
defines the $r$-th Chern character of $SO(N)$. Now, since the curvature
two-form {\bf F} in the vectorial representation is antisymmetric in its
indices, the exponents $\{r_j\}$ are necessarily even, and therefore (\ref
{chern}) vanishes unless $D$ is a multiple of four. Thus, one arrives at the
following lemmas:

{\bf Lemma: 1} For $D=4k$, the only D-forms built from $e^a$, $R^{ab}$ and $%
T^a$, invariant under the AdS group, are the Chern characters for $SO(D+1)$.

{\bf Lemma: 2} For $D=4k+2$, there are no AdS-invariant D-forms constructed
from $e^a$, $R^{ab}$ and $T^a$.

In view of this, it is clear why attempts to construct gravitation theories
with local AdS invariance in even dimensions have been unsuccessful \cite
{freund,M-M}.

Since the forms $P_{r_1 \cdots r_s}$ are closed, they are at best boundary
terms in $4k$ dimensions --which do not contribute to the classical
equations, but could assign different weights to configurations with
nontrivial torsion in the quantum theory. In other words, they can be
locally expressed as 
\begin{equation}
P_{r_1 \cdots r_s} = dL^{AdS}_{\{r\}4k-1}(W).  \label{ads-cs}
\end{equation}
Thus, for each collection $\{r\}$, the ($4k-1$)- form $L^{AdS}_{\{r\}4k-1}$
defines a Lagrangian for the AdS group in $4k-1$ dimensions. It takes direct
computation to see that these Lagrangians involve torsion explicitly. These
results are summarized in the following \newline

{\bf Theorem:} There are two families of gravitational first order
Lagrangians for $e$ and $\omega$, invariant under local AdS transformations:

{\bf a:} Euler-Chern-Simons form in $D=2n-1$, whose exterior derivative is
the Euler character in dimension $2n$, which do not involve torsion
explicitly, and \newline

{\bf b:} Pontryagin-Chern-Simons forms in $D=4k-1$, whose exterior
derivatives are the Chern characters in $4k$ dimensions, which involves
torsion explicitly.

It must be stressed that locally AdS-invariant gravity theories only exist
in odd dimensions. They are $genuine$ gauge systems, whose action comes from
topological invariants in one dimension above. These topological invariants
can be written as the trace of a homogeneous polynomial of degree $n$ in the
AdS curvature. Obviously, for dimensions $4k-1$ both {\bf a}- and {\bf b}%
-families exist. The most general Lagrangian of this sort is a linear
combination of the two families.

An important difference between these two families is that under a parity
transformation the first is even while the second is odd \footnote{%
Parity is understood here as an inversion of one coordinate, both in the
tangent space and in the base manifold. Thus, for instance the Euler
character is invariant under parity, while the Lorentz Chern characters and
the torsional terms are parity violating.}. The parity invariant family has
been extensively studied in \cite{chamslett,chamseddine,btz,jjg}. In what
follows we concentrate on the construction of the pure gravity sector as a
gauge theory which is parity-odd. This construction was discussed in \cite
{tronco}, and also briefly in \cite{trz,trz'}.


\subsection{Even dimensions}


In $D=4$, the the only local Lorentz-invariant 4-forms constructed with the
recipe just described are \cite{mz}: 
\begin{eqnarray*}
{\cal E}_4 & = & \epsilon_{abcd} R^{ab} R^{cd} \\
L_{EH} & = & \epsilon_{abcd} R^{ab} e^c e^d \\
L_C & = & \epsilon_{abcd} e^a e^b e^c e^d \\
C_2 & = & R^{ab} R_{cd} \\
L_{T_1} & = & R^{ab} e_a e_b \\
L_{T_2} & = & T^a T_a. \\
\end{eqnarray*}

The first three terms are even under parity and the rest are odd. Of these, $%
{\cal E}_4$ and $C_2$ are topological invariant densities (closed forms):
the Euler character and the second Chern character for $SO(4)$,respectively.
The remaining four terms define the most general gravity action in four
dimensions, 
\begin{equation}
I= \int_{M_{4}}\left[\alpha L_{EH} +\beta L_C + \gamma L_{T1} + \rho L_{T2}
\right].
\end{equation}
It can also be seen, that by choosing $\gamma=-\rho$, the last two terms are
combined into a topological invariant density (the Nieh-Yan form). Thus,
with this choice the odd part of the action becomes a boundary term.
Furthermore, $C_2$, $L_{T_1}$ and $L_{T_2}$ can be combined into the second
Chern character of the AdS group,

\begin{equation}
R_{\;b}^aR_{\;a}^b + 2(T^a T_a - 2R^{ab}e_a e_b) = R_{\;B}^A R_{\;A}^B.
\end{equation}

This is the only AdS invariant constructed with $e^a$, $\omega^{ab}$ and
their exterior derivatives alone, confirming that there are no locally AdS
invariant gravities in four dimensions.

In general, the only AdS-invariant functionals in higher dimensions can be
written in terms of the AdS curvature as \cite{mz}

\begin{equation}
\tilde{I}_{r_1 \cdots r_s}=\int_{M}C_{r_1} \cdots C_{r_s},
\end{equation}
or linear combinations thereof, where $C_{r}=Tr[(R_{B}^{A})^{r}]$ is the $r$%
-th Chern character for the AdS group. For example, en $D=8$ the Chern
characters for the AdS group are 
\begin{equation}
\begin{array}{c}
Tr[(R_{\;B}^A)^4]=C_4, \\ 
\\ 
Tr[(R_{\;B}^A)^2]\mbox{\tiny $\wedge$}Tr[(R_{\;B}^A)^2]=(C_2)^2.
\end{array}
\label{T7}
\end{equation}
Similar Chern classes are also found for $D=4k$. (As already mentioned, $%
\tilde{I}_{r_1 \cdots r_s}$ vanishes if one of the $r$'s is odd, which is
the case in $4k+2$ dimensions.)

Thus, there are no AdS-invariant gauge theories in even dimensions.


\subsection{Odd dimensions}


The simplest example is found in three spacetime dimensions where there are
two locally AdS-invariant Lagrangians, namely, the Einstein-Hilbert with
cosmological constant, 
\begin{equation}
L_{G\;3}^{AdS}=\epsilon_{abc}[R^{ab}e^c +\frac{1}{3l^2}e^a e^b e^c],
\label{AdS3G}
\end{equation}
and the ``exotic'' Lagrangian 
\begin{equation}
L_{T\;3}^{AdS}= L_3^{*}(\omega) + 2e_{a}T^{a},  \label{LT3}
\end{equation}
where 
\begin{equation}
L_3^{*} \equiv \omega _{b}^{a}d\omega _{a}^{b}+\frac{2}{3}\omega_{b}^{a}
\omega_{c}^{b} \omega_{a}^{c},  \label{L*3}
\end{equation}
is the Lorentz Chern-Simons form. Note that in (\ref{LT3}), the local AdS
symmetry fixes the relative coefficient between $L_3^{*}(\omega)$, and the
torsion term $e_{a}T^{a}$. The most general action for gravitation in $D=3$,
which is invariant under $SO(4)$ is therefore a linear combination $\alpha
L_{G\;3}^{AdS}+ \beta L_{T\;3}^{AdS}$.

For $D=4k-1$, the number of possible exotic forms grows as the partitions of 
$k$, in correspondence with the number of composite Chern invariants of the
form $P_{\{r\}}=\prod_{j}C_{r_j}$. The most general Lagrangian in $4k-1$
dimensions takes the form $\alpha L_{G\;4k-1}^{AdS} + \beta_{\{r\}}
L_{T\;\{r\}\;4k-1}^{AdS}$, where $d L_{T\;\{r\}\;4k-1}^{AdS} = P_{\{r\}}$,
with $\sum_{j}r_j=4k$. These Lagrangians have proper dynamics and, unlike
the even dimensional cases, they are not boundary terms. For example, in
seven dimensions one finds \cite{tronco,cz}\newline
\begin{eqnarray*}
&& L_{T\;7}^{AdS}=\beta_{2,2}[R^a\,_bR^b\,_a + 2(T^a T_a -R^{ab}e_a e_{b})]
L_{T\;3}^{AdS} \\
&& +\beta_4[L_7^{*}(\omega) + 2(T^aT_a +R^{ab}e_ae_b)T^ae_a +
4T_aR^a\,_bR^b\,_ce^c],
\end{eqnarray*}
where $L_{2n-1}^{*}$ is the Lorentz-CS (2n-1)-form, 
\begin{equation}
dL_{2n-1}^{*}(\omega) = Tr[(R^a_{\;b})^n].  \label{7D}
\end{equation}

{\bf Summarizing:} The requirement of local AdS symmetry is rather strong
and has the following consequences:

\begin{itemize}
\item  Locally AdS invariant theories of gravity exist in odd dimensions
only.

\item  For $D=4k-1$ there are two families: one involving only the curvature
and the vielbein (Euler Chern-Simons form), and the other involving torsion
explicitly in the Lagrangian. These families are even and odd under space
reflections, respectively.

\item  For $D=4k+1$ only the Euler-Chern-Simons forms exist. These ar parity
even and don't involve torsion explicitly.
\end{itemize}


\section{Exact Solutions}


As stressed here, the local symmetry of odd-dimensional gravity can be
extended from Lorentz to AdS by an appropriate choice of the free
coefficients in the action. The resulting Lagrangians (with or without
torsion terms), are Chern-Simons $D$-forms defined in terms of the AdS
connection {\bf A}, whose components include the vielbein and the spin
connection [see eqn. (\ref{w})]. This implies that the field equations (\ref
{E-L},\ref{Tor}) obtained by varying the vielbein and the spin connection
respectively, can be written in an AdS-covariant form 
\begin{equation}
<{\bf F}^{n-1}J_{AB}>=0,  \label{eqAdS}
\end{equation}
where {\bf F}$=\frac{1}{2}R^{AB}J_{AB}$ is the AdS curvature with $R^{AB}$
given by (\ref{R}) and $J_{AB}$ are the AdS generators.

It is easily checked that any locally AdS spacetime is a solution of (\ref
{eqAdS}). Apart from anti-de Sitter space itself, some interesting
spacetimes with this feature are the topological black holes of Ref. \cite
{pmb}, and some ``black branes" with constant curvature worldsheet \cite
{AOTZ}. For any $D$, there is also a unique static, spherically symmetric,
asymptotically AdS black hole solution \cite{btz}, as well as their
topological extensions which have nontrivial event horizons \cite{cai}.

Exact solutions of the form AdS$_{4}\times S^{D-4}$ have also been found 
\cite{MTZ} \footnote{%
The de-Sitter case ($\Lambda >0$) was discussed in \cite{Muller} for the
torsion-free theory. Changing the sign in the cosmological constant has deep
consequences. In fact, the solutions are radically different, and locally
supersymmetric extensions for positive cosmological constant don't exist in
general.} as well as alternative four-dimensional cosmological models.

All of the above geometries can be extended into solutions of the
gravitational Born-Infeld theory (\ref{BI'}) in even dimensions.
Friedmann-Robertson-Walker like cosmologies have been shown to exist in even
dimensions \cite{jjg}, and it could be expected that similar solutions
exists in odd dimensions as well.


\section{Chern-Simons Supergravities}


We now consider the supersymmetric extensions of the locally AdS theories
defined above. The idea is to enlarge the AdS algebra incorporating SUSY
generators. The closure of the algebra (Jacobi identity) forces the addition
of further bososnic generators as well \cite{vV}. In order to accomodate
spinors in a natural way, it is useful to cast the AdS generators in the
spinor representation of $SO(D+1)$. In particular, one can write, 
\begin{equation}
dL^{AdS}_{T\;4k-1}= \frac{-1}{2^{4k}}Tr[(R^{AB} \Gamma_{AB})^{2k}].
\label{LT}
\end{equation}
which is a particular form of (\ref{F^n}) where $\left\langle \right\rangle$
has been replaced by the ordinary trace over spinor indices in this
representation.

Other possibilities of the form $\left\langle \mbox{{\bf F}}^{n-p}\right
\rangle \left\langle\mbox{{\bf F}}^p \right\rangle$, are not necessary to
reproduce the minimal supersymmetric extensions of AdS containing the
Hilbert action. In the supergravity theories discussed below, the
gravitational sector is given by $\pm\frac{1}{2^n} L_{G\;2n-1}^{AdS}- \frac{1%
}{2}L_{T\;2n-1}^{AdS}$. The $\pm $ sign corresponds to the two choices of
inequivalent representations of $\Gamma $'s, which in turn reflect the two
chiral representations in $D+1$. As in the three-dimensional case, the
supersymmetric extensions of $L_{G}$ or any of the exotic Lagrangians such
as $L_{T}$, require using both chiralities, thus doubling the algebras. Here
we choose the + sign, which gives the minimal superextension \cite{tronco}.

The bosonic theory (\ref{LT}) is our starting point. The idea now is to
construct its supersymmetric extension. For this, we need to express the
adjoint representation in terms of the Dirac matrices of the appropriate
dimension. This is always possible because the generators of the Dirac
algebra, $\{I$, $\Gamma ^{a}$, $\Gamma ^{ab}$,...$\}$, provide a basis for
the space of square matrices. The advantage of this approach is that it
gives an explicit representation of the algebra and writing the Lagrangians
is straightforward.

The supersymmetric extensions of the AdS algebras in $D=$ 2, 3, 4, mod 8,
were studied by van Holten and Van Proeyen in \cite{vV}. They added one
Majorana supersymmetry generator to the AdS algebra and found all the $N=1$
extensions demanding closure of the full superalgebra. In spite of the fact
that the algebra for $N=1$ AdS supergravity in eleven dimensions was
conjectured in 1978 to be $osp(32|1)$ by Cremer, Julia and Scherk \cite{CJS}%
, and this was confirmed in \cite{vV}, nobody constructed a supergravity
action for this algebra in the intervening twenty years.

One reason for the lack of interest in the problem might have been the fact
that the $osp(32|1)$ algebra contains generators which are Lorentz tensors
of rank higher than two.In the past, supergravity algebras were
traditionally limited to generators which are Lorentz tensors up to second
rank. This constraint was based on the observation that elementary particle
states of spin higher than two would be inconsistent \cite{nahm}. However,
this does not rule out the relevance of those tensor generators in theories
of extended objects \cite{AGIT}. In fact, it is quite common nowadays to
find algebras like the $M-$brane superalgebra \cite{townsend,NG}, 
\begin{equation}
\{Q^{,}\bar{Q}\}\sim \Gamma ^{a}P_{a}+ \Gamma ^{ab}Z_{ab}+ \Gamma^{abcde}
Z_{abcde}.  \label{malgebra}
\end{equation}


\subsection{Superalgebra and Connection}


The smallest superalgebra containing the AdS algebra in the bosonic sector
is found following the same approach as in \cite{vV}, but lifting the
restriction of $N=1$ \cite{tronco}. The result, for odd $D>3$ is (see \cite
{trz'} for details) \newline

\begin{center}
\begin{tabular}{|l|c|c|c|}
\hline
D & S-Algebra & Conjugation Matrix & Internal Metric \\ \hline
$8k-1$ & $osp(N|m)$ & $C^{T}=C$ & $u^{T}=-u$ \\ \hline
$8k+3$ & $osp(m|N)$ & $C^{T}=-C$ & $u^{T}=u$ \\ \hline
$4k+1$ & $su(m|N)$ & $C^{\dag }=C$ & $u^{\dag }=u$ \\ \hline
\end{tabular}
\end{center}

In each of these cases, $m=2^{[D/2]}$ and the connection takes the form 
\begin{eqnarray}
\mbox{{\bf A}}&=& \frac{1}{2}\omega ^{ab}J_{ab}+ e^{a}J_{a} + \frac{1}{r!}
b^{[r]}Z_{[r]}+  \nonumber \\
&&\frac{1}{2}(\bar{\psi}^i Q_i -\bar{Q}^i \psi_i) + \frac{1}{2}a_{ij}M^{ij}.
\label{A}
\end{eqnarray}

The generators $J_{ab},J_{a}$ span the AdS algebra and the $Q_{\alpha}^i$'s
generate (extended) supersymmetry transformations. The $Q$'s transform in a
vector representation under the action of $M_{ij}$ and as spinors under the
Lorentz group. Finally, the $Z$'s complete the extension of AdS into the
larger algebras $so(m)$, $sp(m)$ or $su(m)$, and $[r]$ denotes a set of $r$
antisymmetrized Lorentz indices.

In (\ref{A}) $\bar{\psi}^i=\psi_j^TCu^{ji}$ ($\bar{\psi}^i=\psi_j^{\dag}
Cu^{ji}$ for $D=4k+1$), where $C$ and $u$ are given in the table above.
These algebras admit $(m+N)\times (m+N)$ matrix representations \cite{freund}%
, where the $J$ and $Z$ have entries in the $m\times m$ block, the $M_{ij}$%
's in the $N\times N$ block, while the fermionic generators $Q$ have entries
in the complementary off-diagonal blocks.

Under a gauge transformation, {\bf A} transforms by $\delta${\bf A}$=
\nabla\lambda$, where $\nabla$ is the covariant derivative for the same
connection {\bf A}. In particular, under a supersymmetry transformation, $%
\lambda =\bar{\epsilon}^iQ_i-\bar{Q}^i\epsilon_i$, and 
\begin{equation}
\delta_{\epsilon}\mbox{{\bf A}}=\left[ 
\begin{array}{cc}
\epsilon^k\bar{\psi}_k- \psi^k\bar{\epsilon}_k & D\epsilon_j \\ 
-D\bar{\epsilon}^i & \bar{\epsilon}^i\psi_j-\bar{\psi}^i\epsilon_j
\end{array}
\right],  \label{delA}
\end{equation}
where $D$ is the covariant derivative on the bosonic connection, $%
D\epsilon_j =(d+\frac{1}{2}[e^a\Gamma_a +\frac{1}{2}\omega^{ab}\Gamma_{ab}+ 
\frac{1}{r!} b^{[r]}\Gamma_{[r]}])\epsilon_j -a_j^i \epsilon_i$.


\subsection{D=5 Supergravity}


In this case, as in every dimension $D=4k+1$, there is no torsional
Lagrangians $L_{T}$ due to the vanishing of the Pontrjagin $4k+2$-forms for
the Riemann cirvature. This fact implies that the local supersymmetric
extension will be of the form $L=L_{G}+\cdot \cdot \cdot $.

As shown in the previous table, the appropriate $AdS$ superalgebra in five
dimensions is $su(2,2|N),$ whose generators are $K,J_{a},J_{ab},Q^{\alpha },%
\bar{Q}_{\beta },M^{ij}$, with $a,b=1,...,5$ and $i,j=1,...,N$. The
connection is {\bf A}$=bK+e^{a}J_{a}+\frac{1}{2}\omega
^{ab}J_{ab}+a_{ij}M^{ij}+\bar{\psi}^{i}Q_{i}-\bar{Q}^{j}\psi _{j}$, so that
in the adjoint representation 
\begin{equation}
\mbox{{\bf A}}=\left[ 
\begin{array}{cc}
\Omega _{\beta }^{\alpha } & \psi _{j}^{\alpha } \\ 
-\bar{\psi}_{\beta }^{i} & A_{j}^{i}
\end{array}
\right] ,  \label{cone5}
\end{equation}
with $\Omega _{\beta }^{\alpha }=\frac{1}{2}(\frac{i}{2}bI+e^{a}\Gamma
_{a}+\omega ^{ab}\Gamma _{ab})_{\beta }^{\alpha }$, $A_{j}^{i}=\frac{i}{N}%
\delta _{j}^{i}b+a_{j}^{i}$, and $\bar{\psi}_{\beta }^{i}=\psi ^{\dagger
\alpha j}G_{\alpha \beta }$. Here $G$ is the Dirac conjugate (e. g., $%
G=i\Gamma ^{0}$). The curvature is 
\begin{equation}
\mbox{{\bf F}}=\left[ 
\begin{array}{cc}
\bar{R}_{\beta }^{\alpha } & D\psi _{j}^{\alpha } \\ 
-D\bar{\psi}_{\beta }^{i} & \bar{F}_{j}^{i}
\end{array}
\right]   \label{curva5}
\end{equation}
where 
\begin{eqnarray}
D\psi _{j}^{\alpha } &=&d\psi _{j}^{\alpha }+\Omega _{\beta }^{\alpha }\psi
_{j}^{\beta }-A_{j}^{i}\psi _{i}^{\alpha },  \nonumber \\
\bar{R}_{\beta }^{\alpha } &=&R_{\beta }^{\alpha }-\psi _{i}^{\alpha }\bar{%
\psi}_{\beta }^{i}, \\
\bar{F}_{j}^{i} &=&F_{j}^{i}-\bar{\psi}_{\beta }^{i}\psi _{j}^{\beta }. 
\nonumber
\end{eqnarray}
Here $F_{j}^{i}=dA_{j}^{i}+A_{k}^{i}A_{j}^{k}+\frac{i}{N}db\delta _{j}^{i}$
is the $su(N)$ curvature, and $R_{\beta }^{\alpha }=d\Omega _{\beta
}^{\alpha }+\Omega _{\sigma }^{\alpha }\Omega _{\beta }^{\sigma }$ is the $%
u(2,2)$ curvature. In terms of the standard $(2n-1)$-dimensional fields, $%
R_{\beta }^{\alpha }$ can be written as 
\begin{equation}
R_{\beta }^{\alpha }=\frac{i}{4}db\delta _{\beta }^{\alpha }+\frac{1}{2}%
\left[ T^{a}\Gamma _{a}+(R^{ab}+e^{a}e^{b})\Gamma _{ab}\right] _{\beta
}^{\alpha }.
\end{equation}

In six dimensions the only invariant form is 
\begin{equation}
P=iStr\left[ \mbox{{\bf F}}^{3}\right] ,
\end{equation}
which in this case reads 
\begin{eqnarray}
P &=&Tr\left[ R^{3}\right] -Tr\left[ F^{3}\right]  \\
&+&3\left[ D\bar{\psi}(\bar{R}+\bar{F})D\psi -\bar{\psi}(R^{2}-F^{2}+[R-F](%
\psi )^{2})\psi \right] ,  \nonumber
\end{eqnarray}
where $(\psi )^{2}=\bar{\psi}\psi $. The resulting five-dimensional C-S
density can de descompossed as a sum a a gravitational part, a $b$-dependent
piece, a $su(N)$ gauge part, and a fermionic term, 
\begin{equation}
L=L_{G}^{AdS}+L_{b}+L_{su(N)}+L_{F},
\end{equation}
with 
\begin{equation}
\begin{array}{l}
L_{G}^{AdS}=\frac{1}{8}\epsilon _{abcde}(R^{ab}R^{cd}e^{e}+\frac{2}{3}%
R^{ab}e^{c}e^{d}e^{e}+\frac{1}{5}e^{a}e^{b}e^{c}e^{d}e^{e}) \\ 
\\ 
L_{b}=-(\frac{1}{N^{2}}-\frac{1}{4^{2}})(db)^{2}b+\frac{3}{4}%
(T^{a}T_{a}-R^{ab}e_{a}e_{b}-\frac{1}{2}R^{ab}R_{ab})b \\ 
+\frac{3}{N}bf_{j}^{i}f_{i}^{j} \\ 
\\ 
L_{su(N)}=-(a_{j}^{i}da_{k}^{j}da_{i}^{k}+\frac{3}{2}%
a_{j}^{i}a_{k}^{j}a_{l}^{k}da_{i}^{l}+\frac{3}{5}%
a_{j}^{i}a_{k}^{j}a_{l}^{k}a_{m}^{l}a_{i}^{m}) \\ 
\\ 
L_{F}=\frac{3}{2}\left[ \bar{\psi}(\bar{R}+\bar{F})D\psi -\frac{1}{2}(\psi
)^{2}(\bar{\psi}D\psi )\right] .
\end{array}
\end{equation}

The action is invariant under local gauge transformations, which contain the
local SUSY transformations 
\begin{equation}
\begin{array}{ccl}
\delta e^{a} & = & -\frac{1}{2}(\overline{\epsilon }^{i}\Gamma ^{a}\psi _{i}-%
\overline{\psi }^{i}\Gamma ^{a}\epsilon _{i}) \\ 
\delta \omega ^{ab} & = & \frac{1}{4}(\overline{\epsilon }^{i}\Gamma
^{ab}\psi _{i}-\overline{\psi }^{i}\Gamma ^{ab}\epsilon _{i}) \\ 
\delta b & = & i(\overline{\epsilon }^{i}\psi _{i}-\overline{\psi }%
^{i}\epsilon _{i}) \\ 
\delta \psi _{i} & = & D\epsilon _{i} \\ 
\delta \overline{\psi }^{i} & = & D\overline{\epsilon }^{i} \\ 
\delta a_{j}^{i} & = & i(\overline{\epsilon }^{i}\psi _{j}-\overline{\psi }%
^{i}\epsilon _{j}).
\end{array}
\label{tr5}
\end{equation}

As in $2+1$ dimensions, the Poincar\'{e} supergravity theory is recovered
contracting the super $AdS$ group. Consider the following rescaling of the
fields 
\begin{equation}
\begin{array}{ccc}
e^a & \rightarrow & \frac{1}{\alpha}e^a \\ 
\omega^{ab} & \rightarrow & \omega^{ab} \\ 
b & \rightarrow & \frac{1}{3\alpha}b \\ 
\psi_i & \rightarrow & \frac{1}{\sqrt{\alpha}}\psi_i \\ 
\overline{\psi}^i & \rightarrow & \frac{1}{\sqrt{\alpha}}\overline{\psi}^i
\\ 
a_j^i & \rightarrow & a_j^i.
\end{array}
\label{rescal5}
\end{equation}
Then, if the gravitational constant is also rescaled as $\kappa \rightarrow
\alpha \kappa$, in the limit $\alpha \rightarrow \infty $ the action becomes
that in \cite{btrz}, plus a $su(N)$ CS form, 
\begin{eqnarray}
I &=& \frac{1}{8}\int[ \epsilon_{abcde}R^{ab}R^{cd}e^e -R^{ab}R_{ab}b- \\
&& 2R^{ab}(\overline{\psi}^i\Gamma_{ab}D\psi_i +D\overline{\psi}^i
\Gamma_{ab}\psi_i) + L_{su(N)}] .  \nonumber  \label{contraction}
\end{eqnarray}

The rescaling (\ref{rescal5}) induces a contraction of the super $AdS$
algebra $su(m|N)$ into [super Poincar\'{e}]$\otimes su(N)$, where the second
factor is an automorfism.


\subsection{D=7 Supergravity}


The smallest AdS superalgebra in seven dimensions is $osp(2|8)$. The
connection (\ref{A}) is {\bf A} =$\frac{1}{2}\omega ^{ab}J_{ab}+e^{a}J_{a}+%
\bar{Q}^{i}\psi _{i}+\frac{1}{2}a_{ij}M^{ij}$, where $M^{ij}$ are the
generators of $sp(2)$. In the representation given above, the bracket $%
\left\langle \ \right\rangle $ is the supertrace and, in terms of the
component fields appearing in the connection, the CS form is 
\begin{eqnarray}
L_{7}^{osp(2|8)}(\mbox{{\bf A}}) &=&2^{-4}L_{G\;7}^{AdS}(\omega ,e)-\frac{1}{%
2}L_{T\;7}^{AdS}(\omega ,e)  \nonumber \\
&&-L_{7}^{*sp(2)}(a)+L_{F}(\psi ,\omega ,e,a).
\end{eqnarray}
Here the fermionic Lagrangian is 
\begin{eqnarray*}
L_{F} &=&4\bar{\psi}^{j}(R^{2}\delta
_{j}^{i}+Rf_{j}^{i}+(f^{2})_{j}^{i})D\psi _{i} \\
&&+4(\bar{\psi}^{i}\psi _{j})[(\bar{\psi}^{j}\psi _{k})(\bar{\psi}^{k}D\psi
_{i})-\bar{\psi}^{j}(R\delta _{i}^{k}+f_{i}^{k})D\psi _{k}] \\
&&-2(\bar{\psi}^{i}D\psi _{j})[\bar{\psi}^{j}(R\delta
_{i}^{k}+f_{i}^{k})\psi _{k}+D\bar{\psi}^{j}D\psi _{i}],
\end{eqnarray*}
where $f_{j}^{i}=da_{j}^{i}+a_{k}^{i}a_{j}^{k}$, and $R=\frac{1}{4}%
(R^{ab}+e^{a}e^{b})\Gamma _{ab}+\frac{1}{2}T^{a}\Gamma _{a}$ are the $sp(2)$
and $so(8)$ curvatures, respectively. The supersymmetry transformations (\ref
{delA}) read \newline

\begin{tabular}{llll}
\hspace{1cm} & $\delta e^a =\frac{1}{2}\bar{\epsilon}^i\Gamma^a \psi_i $ & 
\hspace{1cm} & $\delta \omega ^{ab}=-\frac{1}{2}\bar{\epsilon}^i\Gamma^{ab}
\psi_i$ \\ 
&  &  &  \\ 
\hspace{1cm} & $\delta \psi_i =D\epsilon_i$ & \hspace{1cm} & $\delta a_j^i = 
\bar{\epsilon}^i \psi_j-\bar{\psi}^i \epsilon_j.$ \label{susy7}
\end{tabular}
\newline

Standard seven-dimensional supergravity is an $N=2$ theory (its maximal
extension is N=4), whose gravitational sector is given by the
Einstein-Hilbert action with cosmological constant and with an $osp(2|8)$
invariant background\cite{D=7,Salam-Sezgin}. In the case presented here, the
extension to larger $N$ is straighforward: the index $i$ is allowed to run
from $2$ to $2s$, and the Lagrangian is a CS form for $osp(2s|8)$.


\subsection{D=11 Supergravity}


In this case, the smallest AdS superalgebra is $osp(32|1)$ and the
connection is {\bf A} =$\frac{1}{2}\omega^{ab}J_{ab} + e^aJ_a + \frac{1}{5!}
b^{abcde}J_{abcde}+ \bar{Q}\psi$, where $b$ is a totally antisymmetric
fifth-rank Lorentz tensor one-form. Now, in terms of the elementary bosonic
and fermionic fields, the CS form in (\ref{F^n}) reads 
\begin{equation}
L_{11}^{osp(32|1)}(${\bf A}$)= L_{11}^{sp(32)}(\Omega )+L_{F}(\Omega,\psi),
\label{L11}
\end{equation}
where $\Omega\equiv \frac{1}{2}(e^{a}\Gamma_{a} + \frac{1}{2}
\omega^{ab}\Gamma_{ab}+ \frac{1}{5!}b^{abcde}\Gamma_{abcde})$ is an $sp(32)$
connection. The bosonic part of (\ref{L11}) can be written as 
\begin{eqnarray}
L_{11}^{sp(32)}(\Omega)&=&2^{-6} L_{G\;11}^{AdS}(\omega,e) -\frac{1}{2}
L_{T\;11}^{AdS}(\omega,e) + L_{11}^b(b,\omega,e).  \nonumber
\end{eqnarray}
The fermionic Lagrangian is 
\begin{eqnarray*}
L_{F} &=&6(\bar{\psi}R^{4}D\psi)-3\left[ (D\bar{\psi}D\psi )+(\bar{\psi}
R\psi)\right] (\bar{\psi}R^{2}D\psi)  \nonumber \\
& &-3\left[(\bar{\psi}R^3\psi)+(D\bar{\psi}R^2 D\psi)\right](\bar{\psi}
D\psi)+ \\
& &2\left[ (D\bar{\psi}D\psi )^{2}+(\bar{\psi}R\psi )^{2}+(\bar{\psi}R\psi)
(D\bar{\psi}D\psi )\right] (\bar{\psi}D\psi),
\end{eqnarray*}
where $R=d\Omega +\Omega ^{2}$ is the $sp(32)$ curvature. The supersymmetry
transformations (\ref{delA}) read \newline
\newline
\begin{tabular}{llll}
\hspace{1cm} & $\delta e^{a}=\frac{1}{8}\bar{\epsilon}\Gamma ^{a}\psi$ & 
\hspace{1cm} & $\delta\omega^{ab}=-\frac{1}{8}\bar{\epsilon}\Gamma^{ab}\psi$
\\ 
&  &  &  \\ 
\hspace{1cm} & $\delta \psi =D\epsilon $ & \hspace{1cm} & $\delta b^{abcde}= 
\frac{1}{8}\bar{\epsilon}\Gamma ^{abcde}\psi.$ \\ 
&  &  &  \\ 
\label{susy11} &  &  & 
\end{tabular}

Standard eleven-dimensional supergravity \cite{CJS} is an N=1 supersymmetric
extension of Einstein-Hilbert gravity that cannot accomodate a cosmological
constant \cite{B-D-H-S,Deser}. An $N>1$ extension of this theory is not
known. In our case, the cosmological constant is necessarily nonzero by
construction and the extension simply requires including an internal $so(N)$
gauge field coupled to the fermions, and the resulting Lagrangian is an $%
osp(32|N)$ CS form \cite{tronco}.


\section{Discussion}


The supergravities presented here have two distinctive features: The
fundamental field is always the connection {\bf A} and, in their simplest
form, they are pure CS systems (matter couplings are discussed below). As a
result, these theories possess a larger gravitational sector, including
propagating spin connection. Contrary to what one could expect, the
geometrical interpretation is quite clear, the field structure is simple
and, in contrast with the standard cases, the supersymmetry transformations
close off shell without auxiliary fields.

{\bf A. Torsion.} It can be observed that the torsion Lagrangians ($L_T$)are
odd while the torsion-free terms ($L_G$) are even under spacetime
reflections. The minimal supersymmetric extension of the AdS group in $4k-1$
dimensions requires using chiral spinors of $SO(4k)$ \cite{Gunaydin}. This
in turn implies that the gravitational action has no definite parity, but
requires the combination of $L_T$ and $L_G$ as described above. In $D=4k+1$
this issue doesn't arise due to the vanishing of the torsion invariants,
allowing constructing a supergravity theory based on $L_G$ only, as in \cite
{chamseddine}. If one tries to exclude torsion terms in $4k-1$ dimensions,
one is forced to allow both chiralities for $SO(4k)$ duplicating the field
content, and the resulting theory has two copies of the same system \cite
{horava}.

{\bf B. Field content and extensions with N$>$1.} The field content compares
with that of the standard supergravities in $D=5,7,11$ as follows:\newline

\begin{center}
\begin{tabular}{c|c|l|l|}
\cline{2-4}
\hspace{.7cm} & D & Standard supergravity & CS supergravity \\ \cline{2-4}
& 5 & $e^a_{\mu}$ $\psi^{\alpha}_{\mu}$ $\bar{\psi}_{\alpha \mu}$ & $%
e_{\mu}^a$ $\omega^{ab}_{\mu}$ $\psi^{\alpha}_{\mu}$ $\bar{\psi}_{\alpha \mu}
$ $b$ \\ \cline{2-4}
& 7 & $e^a_{\mu}$ $A_{[3]}$ $\psi^{\alpha i}_{\mu}$ $a^i_{\mu j}$ $%
\lambda^{\alpha}$ $\phi$ & $e_{\mu }^a$ $\omega ^{ab}_{\mu}$ $%
\psi_{\mu}^{\alpha i }$ $a_{\mu j}^i$ \\ \cline{2-4}
& 11 & $e^a_{\mu}$ $A_{[3]}$ $\psi^{\alpha}_{\mu}$ & $e_{\mu}^{a}$ $\omega
^{ab}_{\mu }$ $\psi_{\mu }^{\alpha }$ $b^{abcde}_{\mu }$ \\ \cline{2-4}
\end{tabular}
\end{center}

Standard supergravity in five dimensions..... The theory obtained with our
scheme is the same one discussed by Chamseddine in \cite{chamseddine}.

Standard seven-dimensional supergravity is an $N=2$ theory (its maximal
extension is N=4), whose gravitational sector is given by Einstein-Hilbert
gravity with cosmological constant and with a background invariant under $%
OSp(2|8)$ \cite{D=7,Salam-Sezgin}. Standard eleven-dimensional supergravity 
\cite{CJS} is an N=1 supersymmetric extension of Einstein-Hilbert gravity
that cannot accomodate a cosmological constant \cite{B-D-H-S,Deser}. An $N>1$
extension of this theory is not known.

In the case presented here, the extensions to larger $N$ are straighforward
in any dimension. In $D=7$, the index $i$ is allowed to run from $2$ to $2s$%
, and the Lagrangian is a CS form for $osp(2s|8)$. In $D=11$, one must
include an internal $so(N)$ field and the Lagrangian is an $osp(32|N)$ CS
form \cite{trz}. The cosmological constant is necessarily nonzero in all
cases.

{\bf C. Spectrum.} The stability and positivity of the energy for the
solutions of these theories is a highly nontrivial problem. As shown in Ref. 
\cite{bgh}, the number of degrees of freedom of bosonic CS systems for $%
D\geq 5$ is not constant throughout phase space and different regions can
have radically different dynamical content. However, in a region where the
rank of the symplectic form is maximal the theory behaves as a normal gauge
system, and this condition is stable under perturbations. As it is shown in 
\cite{CTZ}, there exists a nontrivial extension of the AdS superalgebra with
one abelian generator for which anti-de Sitter space without matter fields
is a background of maximal rank, and the gauge superalgebra is realized in
the Dirac brackets. For example, for $D=11$ and $N=32$, the only
nonvanishing anticommutator reads 
\begin{eqnarray*}
\{Q^i_{\alpha},\bar{Q}^j_{\beta} \} &=& \frac{1}{8}\delta^{ij}\left[
C\Gamma^{a} J_a + C\Gamma^{ab}J_{ab} + C\Gamma^{abcde}Z_{abcde}
\right]_{\alpha \beta} \\
& & -M^{ij}C_{\alpha \beta},
\end{eqnarray*}
where $M^{ij}$ are the generators of $SO(32)$ internal group. On this
background the $D=11$ theory has $2^{12}$ fermionic and $2^{12}-1$ bosonic
degrees of freedom. The (super)charges obey the same algebra with a central
extension. This fact ensures a lower bound for the mass as a function of the
other bosonic charges \cite{G-H}.

{\bf D. Classical solutions.} The field equations for these theories in
terms of the Lorentz components ($\omega$, $e$, $b$, $a$, $\psi$) are
spread-out expressions for $<${\bf F}$^{n-1}G_{(a)}> =0$, where $G_{(a)}$
are the generators of the superalgebra. It is rather easy to verify that in
all these theories the anti-de Sitter space is a classical solution , and
that for $\psi=b=a=0$ there exist spherically symmetric, asymptotically AdS
standard \cite{jjg}, as well as topological \cite{pmb} black holes. In the
extreme case these black holes can be shown to be BPS states.

{\bf E. Matter couplings.} It is possible to introduce a minimal couplings
to matter of the form {\bf A}$\cdot ${\bf J}. For $D=11$, the matter content
is that of a theory with (super-) 0, 2, and 5--branes, whose respective
worldhistories couple to the spin connection and the $b$ fields.

{\bf F. Standard SUGRA.} Some sector of these theories might be related to
the standard supergravities if one identifies the totally antisymmetric part
of the contorsion tensor in a coordinate basis, $k_{\mu \nu \lambda}$, with
the abelian 3-form, $A_{[3]}$. In 11 dimensions one could also identify the
antisymmetric part of $b$ with an abelian 6-form $A_{[6]}$, whose exterior
derivative, $dA_{[6]}$, is the dual of $F_{[4]}=dA_{[3]}$. Hence, in $D=11$
the CS theory possibly contains the standard supergravity as well as some
kind of dual version of it.\newline

{\bf ACKNOWLEDGEMENTS} \newline

The authors are grateful to R. Aros, M. Ba\~nados, O. Chand\'{\i}a, M.
Contreras, A. Dabholkar, S. Deser, G. Gibbons, A. Gomberoff, M. G\"unaydin,
M. Henneaux, C. Mart\'{\i}nez, F. M\'endez, S. Mukhi, R. Olea, C. Teitelboim
and E. Witten for many enlightening discussions and helpful comments. This
work was supported in part by grants 1960229, 1970151, 1980788 and 3960009
from FONDECYT (Chile), and 27-953/ZI-DICYT (USACH). Institutional support to
CECS from Fuerza A\'{e}rea de Chile and a group of Chilean private companies
(Business Design Associates, CGE, CODELCO, COPEC, Empresas CMPC, Minera
Collahuasi, Minera Escondida, NOVAGAS and XEROX-Chile) is also acknowledged.


\begin{references}
\bibitem{}  

\bibitem{trz}  R. Troncoso and J. Zanelli, Phys. Rev. {\bf D 58} R101703,
(1998). 

\bibitem{trz'}  R. Troncoso and J. Zanelli,{\em Gauge Supergravities for all
Odd Dimensions}, lecture presented at the Third Meeting Quantum Gravity in
the Southern Cone, Bariloche, January 1998. hep-th/9807029 

\bibitem{MH}  M. Henneaux, Phys. Rep. {\bf 126}(1985)1. 

\bibitem{bgh}  M. Ba\~{n}ados, L. J. Garay and M. Henneaux, Phys. Rev.{\bf %
D53} R593 (1996); Nucl. Phys.{\bf B476} 611 (1996). 

\bibitem{lanczos}  C. Lanczos, Ann. Math. {\bf 39} (1938) 842. 

\bibitem{lovelock}  D. Lovelock, J. Math. Phys. {\bf 12} (1971) 498. 

\bibitem{zumino}  B. Zumino, Phys. Rep. {\bf 137} (1986) 109. 

\bibitem{mz}  A. Mardones and J. Zanelli, Class. Quantum Grav. {\bf 8}(1991)
1545. 

\bibitem{pkt84}  P. K. Townsend, {\em Three Lectures on Quantum
Supersymmetry and Supergravity}, Trieste Summer School '84, B. de Wit, P.
Fayet, and P. van Nieuwenhuizen, editors. 

\bibitem{tr}  J. C. Taylor and V. O. Rivelles, Phys. Lett. {\bf B104} (1981)
131; {\bf B121} (1983) 38. 

\bibitem{witten}  E. Witten, Nucl. Phys. {\bf B311} (1988) 46. 

\bibitem{chamslett}  A.Chamseddine, Phys. Lett. {\bf B233} (1989) 291. 

\bibitem{chamseddine}  A.Chamseddine, Nucl.Phys. {\bf B346} (1990) 213. 

\bibitem{btz}  M. Ba\~{n}ados, C. Teitelboim and J. Zanelli, Phys. Rev. {\bf %
D49} (1994) 975. 

\bibitem{z}  J. Zanelli, Phys. Rev.{\bf D51} (1995) 490. 

\bibitem{btrz}  M. Ba\~{n}ados, R. Troncoso and J. Zanelli, Phys. Rev. {\bf %
D54} (1996) 2605. 

\bibitem{pvn}  P. van Nieuwenhuizen, Phys. Rep. {\bf 68} (1981) 1. 

\bibitem{vV}  J. W. van Holten and A. Van Proeyen, J. Phys. {\bf A 15}(1982)
3763. 

\bibitem{soh}  M. Sohnius, Phys. Rep. {\bf 28} (1985) 39. 

\bibitem{Gunaydin}  M. G\"{u}naydin and C. Saclioglu, Comm. Math. Phys. {\bf %
87} (1982) 159. 

\bibitem{Regge}  T.Regge, Phys.Rep. {\bf 137},(1986) 31. 

\bibitem{tz}  C. Teitelboim and J. Zanelli, Class. and Quantum Grav. {\bf 4}%
(1987) L125. 

\bibitem{zwiebach}  B. Zwiebach, Phys. Lett. {\bf 156B} (1985) 315. 

\bibitem{bd}  D. G. Boulware and S. Deser, Phys. Rev. Lett. {\bf 55} (1985)
2656. 

\bibitem{jjg}  M. Ba\~{n}ados, C. Teitelboim and J. Zanelli, {\em Lovelock-
Born-Infeld Theory of Gravity} in {\em J. J. Giambiagi Festschrift}, H.
Falomir, E.Gamboa-Sarav\'{\i}, P. Leal, and F. Schaposnik (eds.),World
Scientific, Singapore, 1991. 

\bibitem{wheeler}  J.T.Wheeler, Nucl. Phys.{\bf B268} (1986) 737; {\bf B273}
(1986) 732. B.Whitt, Phys. Rev. {\bf D38} (1988) 3001. R. C. Myers and J.
Simon, Phys. Rev. {\bf D38} (1988) 2434. D.L.Wiltshire, {\it ibid.}, {\bf 38}
(1988) 2445. 

\bibitem{htz}  M. Henneaux, C. Teitelboim and J. Zanelli, {\em Gravity in
Higher Dimensions}, in SILARG V, M. Novello, (ed.), World Scientific,
Singapore, 1987; Phys. Rev. {\bf A 36} (1987) 4417. 

\bibitem{hojman}  R.Hojman, C.Mukku and W.A.Sayed, Phys.Rev. {\bf D22}
(1980) 1915. 

\bibitem{mauricio}  M.Contreras and J.Zanelli, {\em A note on the spin
connection formulation of gravity} (to appear). 

\bibitem{freund}  P.Freund, {\em Introduction to Supersymmetry} Cambridge
University Press, Cambridge, U.K., 1989. 

\bibitem{E-C}  R. Debever, (ed.), {\em Elie Cartan -- Albert Einstein,
Lettres sur le Parall\'{e}lisme Absolu, 1929-1932} Acad\'{e}mie Royal de
Belgique \& Princeton University Press (1979). 

\bibitem{nakahara}  M. Nakahara,{\em Geometry, Topology and Physics} Adam
Hilger, New York, 1990. T. Eguchi, P.B.Gilkey, and A.J.Hanson, Phys. Rep. 
{\bf 66} (1980) 213. 

\bibitem{M-M}  S.W.MacDowell and F.Mansouri, Phys.Rev.Lett.{\bf 38} (1977)
739; Erratum-ibid.{\bf 38} (1977) 1376. 

\bibitem{tronco}  R. Troncoso, Doctoral Thesis, University of Chile (1996). 

\bibitem{cz}  O.Chand\'{\i }a and J.Zanelli, {\em Torsional Topological
Invariants (and their relevance for Real Life)}. Lecture given at La Plata
Meeting on Trends in Theoretical Physics, La Plata, Argentina, May 1997,
hep-th/9708138. 

\bibitem{pmb}  S. Aminneborg, I. Bengtsson, S. Holst and P. Peldan, Class.
Quantum Grav.{\bf 13} (1996) 2707. M. Ba\~{n}ados,Phys. Rev. {\bf D57}
(1998), 1068. R.B. Mann, {\em Topological Black Holes: Outside Looking In},
gr-qc/9709039. 

\bibitem{AOTZ}  R. Aros, R. Olea, R. Troncoso and J. Zanelli, {\em Constant
Curvature Black Branes} (manuscript in preparation). 

\bibitem{cai}  R.Cai and K.Soh, {\em Topological Black Holes in
Dimensionally Continued Gravity}, gr-qc/9808067 

\bibitem{MTZ}  C.Mart\'{\i}nez, R.Troncoso and J.Zanelli (manuscript in
preparation). 

\bibitem{Muller}  F. M\"uller-Hoissen, Nucl.Phys.B346:235-252,1990 

\bibitem{CJS}  E. Cremmer, B. Julia and J. Scherk, Phys. Lett. {\bf 76B}%
(1978) 409. 

\bibitem{nahm}  W. Nahm, Nucl. Phys. {\bf B135} (1978) 149. J. Strathdee,
Int. J. Mod. Phys. {\bf A 2} (1987) 273. 

\bibitem{AGIT}  J. A. de Azc\'{a}rraga, J. P. Gauntlet, J. M. Izquierdo and
P. K. Townsend, Phys. Rev. Lett. {\bf 63} (1989) 2443. 

\bibitem{townsend}  P. K. Townsend, {\em p-Brane Democracy}, hep-th/9507048 

\bibitem{NG}  H. Nishino and S. J. Gates, Phys. Lett. {\bf B388} (1996) 504. 

\bibitem{D=7}  P.K.Townsend and P.van Nieuwenhuizen, Phys. Lett. {\bf 125B}
(1983) 41. 

\bibitem{Salam-Sezgin}  A.Salam and E.Sezgin, Phys. Lett. {\bf 126B} (1983)
295. 

\bibitem{B-D-H-S}  K.Bautier, S.Deser, M.Henneaux and D.Seminara, Phys.
Lett. {\bf B406},(1997) 49. 

\bibitem{Deser}  S.Deser, {\em Uniqueness of D=11 Supergravity}, Lecture
presented at this meeting, August 1997, hep-th/9712064. 

\bibitem{CTZ}  O.Chand\'{\i }a, R. Troncoso and J.Zanelli, (in preparation). 

\bibitem{G-H}  G. W. Gibbons and C. M. Hull, Phys. Lett. {\bf 109B}, 190
(1982). 

\bibitem{horava}  P.Horava, {\em M-Theory as a Holographic Field Theory}
hep-th/9712130.
\end{references}
\end{document}